\documentclass[
    ,final            
  ]
  {aipproc}

\layoutstyle{6x9}

\begin{document}

\title{Perturbative Renormalizability of Chiral Two Pion Exchange
and Power Counting in Nucleon-Nucleon Scattering}



\author{Manuel Pav\'on Valderrama}{
  address={Departamento de F\'{\i}sica Te\'orica and
Instituto de F\'{\i}sica Corpuscular (IFIC),
Centro Mixto CSIC-Universidad de Valencia,
Institutos de Investigaci\'on de Paterna, Aptd. 22085, E-46071 Valencia, Spain},
email={m.pavon.valderrama@ific.uv.es}
}

\begin{abstract}
We show how to renormalize chiral two pion exchange perturbatively
if one pion exchange has already been fully iterated at leading order.
This particular choice corresponds to the implementation of
the counting proposal of Nogga, Timmermans and van Kolck
at subleading orders.
We illustrate why the perturbative treatment of the two pion exchange
contributions is mandatory in order to avoid certain inconsistencies
in Weinberg's counting.
In addition, renormalizability implies modifications of the power counting
which we explore for the particular case of the singlet channel.
\end{abstract}

\maketitle


\section{Introduction}

The effective field theory (EFT) formulation of nuclear forces~\cite{Beane:2000fx,Bedaque:2002mn,Epelbaum:2005pn,Epelbaum:2008ga}
tries to provide a consistent understanding of nuclear physics in terms
of chiral symmetry, the main low energy manifestation
of quantum chromodynamics.
The applicability of EFT techniques relies on the existence of a separation
of scales in the nuclear force:
the long distance physics is known to be dominated
by pion exchanges (see for example~\cite{Rentmeester:1999vw}
for a demonstration), which in turn are
constrained by the requirements of chiral symmetry.
On the contrary, the nature of the interaction at short distances is poorly
understood, and has been traditionally treated in a purely
phenomenological manner.
However, the specific parametrization used for the short range physics
is inessential for the description of low energy phenomena.

In Weinberg's original formulation~\cite{Weinberg:1990rz,Weinberg:1991um},
which represents the first proposal for constructing an EFT of nuclear forces,
the nuclear potential is expanded as a power series (or power counting)
in terms of the low energy scales of the system,
such as the pion mass or the nucleon momentum.
The resulting chiral potential is inserted into the Schr\"odinger or
Lippmann-Schwinger equation,
from which wave functions and observables can be computed.
This prescription takes into account the non-perturbative nature
of nuclear forces and fits naturally into the traditional paradigm
of nuclear physics.
The Weinberg approach is phenomenologically very successful, as exemplified
by the ${\rm N^3LO}$ calculations of Refs.~\cite{Entem:2003ft,Epelbaum:2004fk}.
From a theoretical point of view, however, the previous calculations are
rather unsatisfactory in the sense that the cut-off is fined tuned
inside a narrow window.
We expect any EFT calculation to be fairly cut-off independent,
a prospect which has encouraged the search for alternatives to Weinberg,
like the KSW counting~\cite{Kaplan:1998tg,Kaplan:1998we,Beane:2008bt},
or, more recently, non-perturbative renormalization~\cite{Nogga:2005hy,PavonValderrama:2005gu,Valderrama:2005wv,PavonValderrama:2007nu,Entem:2007jg,PavonValderrama:2010fb}.
Although the requirement of renormalizability has been put in question
in~\cite{Epelbaum:2006pt,Epelbaum:2009sd}
(see also Ref.~\cite{Machleidt:2010kb} for a balanced discussion on the merits
and disadvantages of the different approaches),
we will show here that cut-off dependence is not the only problem
that affects the Weinberg scheme.

In this contribution we will implement the power counting proposal
of Nogga, Timmermans and van Kolck (NTvK)~\cite{Nogga:2005hy}
at next-to-leading and
next-to-next-to-leading order~\cite{Valderrama:2009ei}.
In this approach, the leading order piece of the chiral nucleon-nucleon (NN)
interaction, one pion exchange, is iterated to all orders and
renormalized, a requisite which implies the modification of
the power counting for certain short range operators.
The subleading pieces of the interaction are treated in perturbation theory,
and as we will show for the particular case of the singlet channel,
cut-off independence induces further modifications
to the power counting~\cite{Valderrama:2009ei}
which were not present in the original Weinberg formulation.
These modifications are compatible with the renormalization group analysis
of the NTvK counting made by Birse~\cite{Birse:2005um}.
We also discuss the role of the cut-off and the possible interpretation
of regularization and renormalization in EFT.

\section{Power Counting}


In Weinberg's power counting~\cite{Weinberg:1990rz,Weinberg:1991um},
the NN potential is described as a low energy expansion
in terms of a ratio of scales, $Q/\Lambda_0$
\begin{eqnarray}
V_{\rm NN}(\vec{q}) = V_{\chi}^{(0)}(\vec{q}) + V_{\chi}^{(2)}(\vec{q}) + 
V_{\chi}^{(3)}(\vec{q}) + {\mathcal O}(\frac{Q^4}{\Lambda_0^4}) \, , 
\end{eqnarray}
where $Q$ represents the low energy scales of the system, such as the pion
mass $m_{\pi}$ or the momentum of the nucleons $p$,
and $\Lambda_0$ represents the high energy scales of the system,
like the nucleon mass $M_N$ or the rho mass $m_{\rho}$.
The rules by which a certain operator or diagram is assigned a given
power of $Q/\Lambda_0$ are called a power counting:
in Weinberg's proposal it is implicitly assumed that
naive dimensional analysis provides a good power counting for the NN potential.
The potential is expected to be iterated at all orders in the Schr\"odinger
or Lippmann-Schwinger equation.

The organization of the chiral potential is as follows:
at order $Q^0$ or leading order (${\rm LO}$) the potential consists
on one pion exchange (OPE) and two contact interactions.
At order $Q^2$ or next-to-leading order (${\rm NLO}$),
leading chiral two pion exchange (TPE) and seven derivative contact terms
are added.
At order $Q^3$ or next-to-next-to-leading order (${\rm N^2LO}$)
subleading two pion exchange enters the potential.
The pion piece of the potential is constrained by chiral symmetry,
while the contact piece is generally used to fit parameters.

\begin{figure}
\includegraphics[height=5.0cm, width=7.0cm]{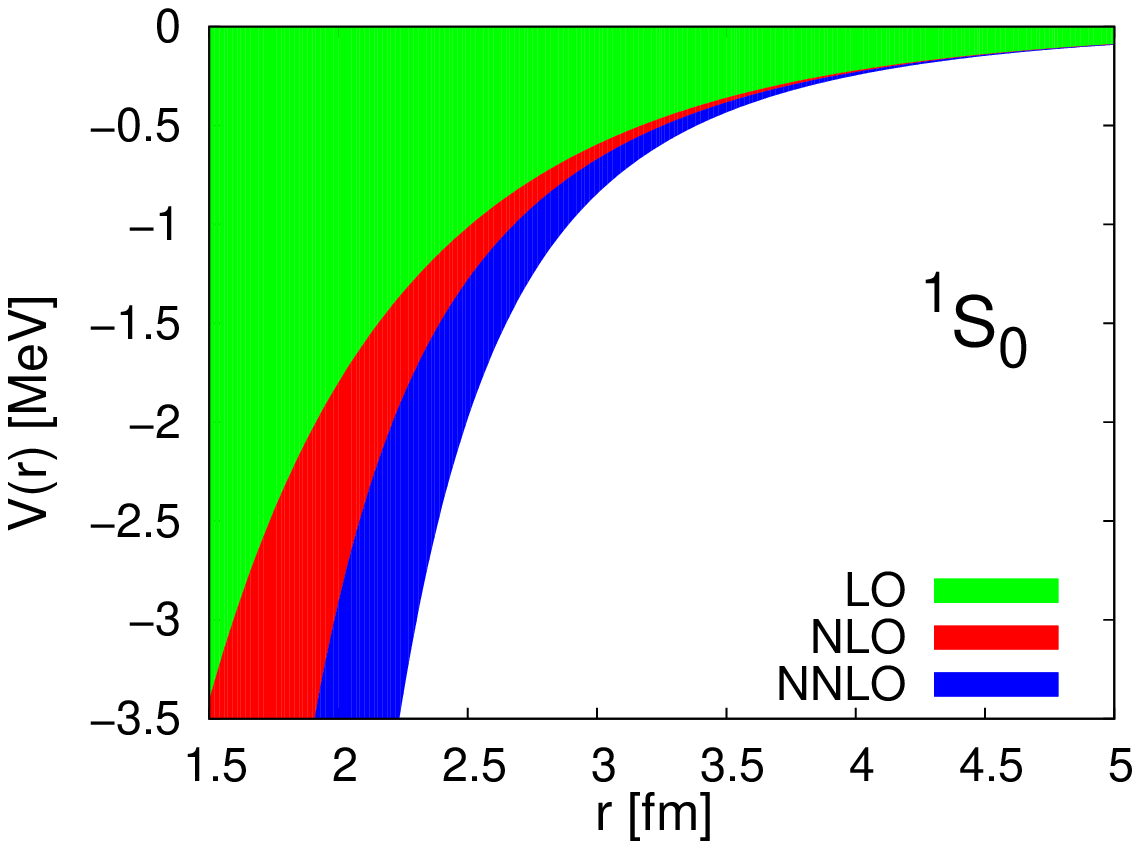}
\includegraphics[height=5.0cm, width=7.0cm]{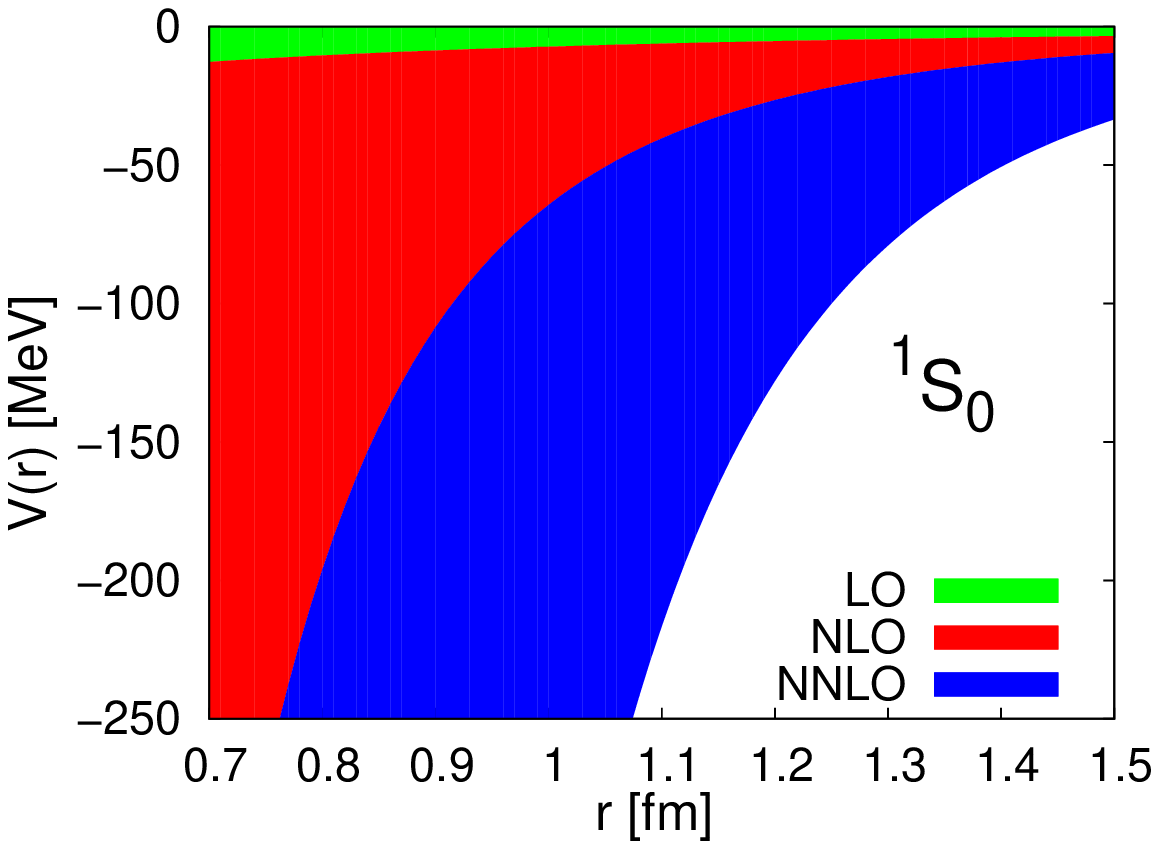}
\caption{
The pion (finite range) component of the chiral nucleon-nucleon potential
in the $^1S_0$ channel.
In the left panel (a) we show the chiral potential
in the $r = 1.5-5.0\,{\rm fm}$ range:
at these distances, each additional contribution to the chiral potential
is suppressed due to power counting.
This results in a convergent pattern for the chiral potentials
at long distances.
In the right panel (b), the different contributions to the chiral potential
in the $r = 0.7-1.5\,{\rm fm}$ range are plotted.
At short distances we see the opposite situation: higher order contributions
are increasingly singular, and the chiral expansion of the nucleon-nucleon
potential does not converge.
}
\label{fig:potentials}
\end{figure}

Power counting determines the convergence of the chiral expansion
of the NN potential:
we expect the chiral potential to converge at large distances / small momenta,
that is, soft scales, and to diverge at hard scales.
The previous is displayed in Fig.~\ref{fig:potentials},
in which we can see how at distances above $2-3\,{\rm fm}$
the leading order of the chiral expansion of the potential dominates,
while the subleading contributions only represent a small correction
to the leading order piece.
In this regard, the chiral NN potentials realize the well known
long distance dominance of one pion exchange
in traditional nuclear physics.
However, as can also be seen in  Fig.~\ref{fig:potentials},
at distances below $1-2\,{\rm fm}$ each additional contribution
to the chiral potential is bigger than the previous one,
meaning that the chiral expansion will not converge.
In fact, on dimensional grounds we expect the pion piece of the order $Q^{\nu}$
contribution to the chiral potential to behave as
\begin{eqnarray}
V^{(\nu)}_{\chi, pions}(\vec{q}) \sim
\frac{|\vec{q}|^{\nu}}{\Lambda_0^{\nu}}\,f(\frac{|\vec{q}|}{m_{\pi}}) \, ,
\end{eqnarray}
where $\vec{q}$ is the momentum exchanged between the nucleons and
$f(x)$ is some non-polynomial function, like a logarithm.
Fourier-transforming the previous expression to coordinate space,
we find
\begin{eqnarray}
V^{(\nu)}_{\chi, pions}(\vec{r}) \sim
\frac{1}{\Lambda_0^{\nu}\,r^{\nu + 3}}\,g(m_{\pi}\,r) \, ,
\end{eqnarray}
where $g(x)$ contains an exponential factor $e^{- n x}$,
with $n$ the number of pions which are exchanged between the nucleons.

The behaviour of the order $Q^{\nu}$ pieces of the chiral potential poses
the problem of how to treat the related short range singularities.
The usual way to deal with this is to apply a renormalization procedure.
The recipe is (i) to regularize the short range pieces of the the potential,
usually by including a cut-off in the calculations, for example
$V(r) \to V(r)\,\theta (r - r_c)$ if we are working in coordinate space
or $\langle p' | V | p \rangle \to \langle p' | V | p \rangle\,\theta(\Lambda - p')\,\theta(\Lambda - p)$ if we are in momentum space, and
(ii) use the counterterms to absorb the undesired cut-off dependence
induced by the previous regularization procedure.
Then we iterate the potential in the Schr\"odinger or Lippmann-Schwinger
equation and fix the finite part of the counterterms to fit observables.
As commented in the introduction, this scheme has led to an impressive
phenomenological description of NN scattering~\cite{Entem:2003ft,Epelbaum:2004fk}.

In the panel (a) of Fig.~\ref{fig:Weinberg-breakdown}
we show a particular realization
of the Weinberg counting at ${\rm N^2LO}$ for the $^1S_0$ singlet channel.
Following~\cite{Epelbaum:2003xx}, we use $f_{\pi} = 92.4\,{\rm MeV}$,
$m_{\pi} = 138.04\,{\rm MeV}$, $d_{18} = -0.97\,{\rm GeV}^2$,
$c_1 = -0.81\,{\rm GeV}^{-1}$, $c_3 = -3.40\,{\rm GeV}^{-1}$
and $c_4 = 3.40\,{\rm GeV}^{-1}$.
We do not implement however spectral regularization.
The calculations are done by solving the Lippmann-Schwinger equation
with a gaussian regulator of the type $e^{-p^6/\Lambda^6}$ and
a cut-off $\Lambda = 400\,{\rm MeV}$.
The counterterms $C_0$ and $C_2$ have been fitted to reproduce
the Nijmegen II phase shifts~\cite{Stoks:1994wp}
for momenta $k \leq 200\,{\rm MeV}$.

\subsection{Inconsistencies in the Weinberg Counting}

There is an inherent problem in iterating the full potential in
order to obtain observables.
As the chiral potentials become increasingly singular at hard scales,
it may be possible for the subleading pieces of the potential
to dominate the amplitudes, even at moderate energies,
if the cut-off is hard enough.
The possibility of a power counting breakdown due to the interplay of
iterations and large cut-offs was already discussed
by Lepage~\cite{Lepage:1997cs} and more recently
by Epelbaum and Gegelia~\cite{Epelbaum:2009sd}.
The problem is whether the cut-offs usually taken in nuclear 
EFT ($\Lambda \sim 0.5\,{\rm GeV}$) are hard enough to trigger
this undesirable situation.

\begin{figure}
\includegraphics[height=4.0cm, width=5.0cm]{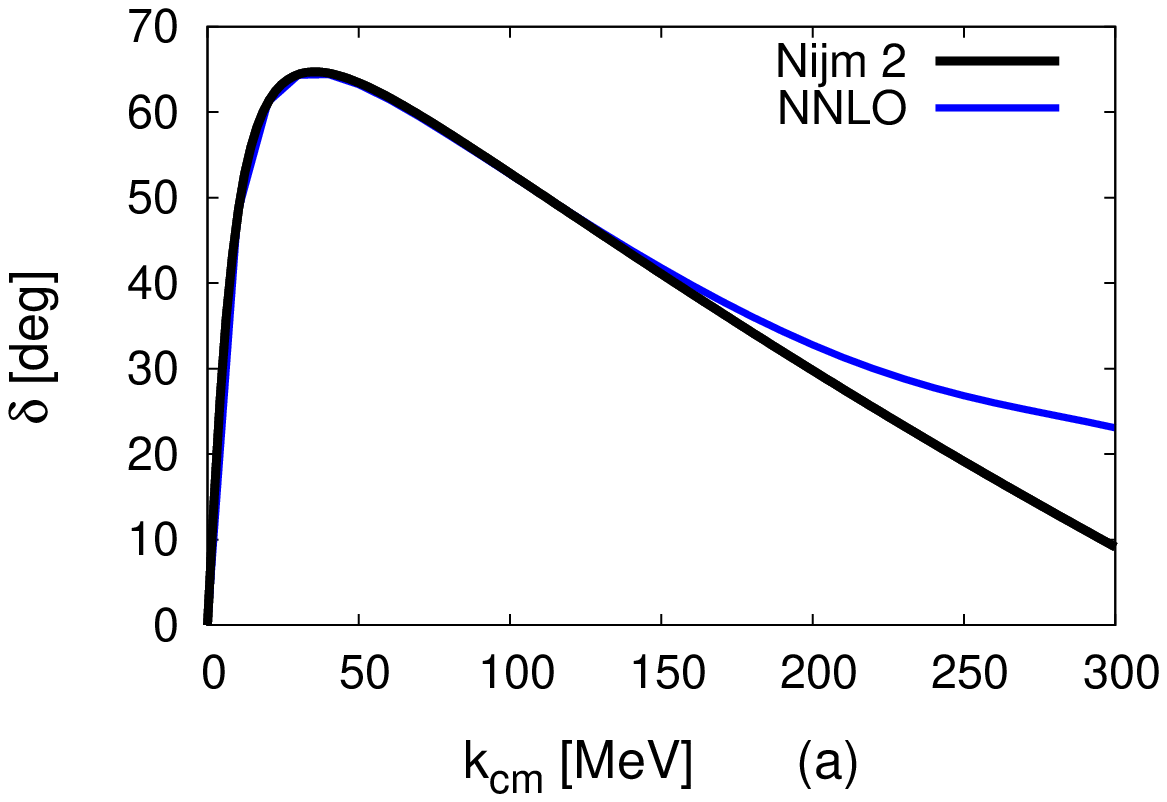}
\includegraphics[height=4.0cm, width=5.0cm]{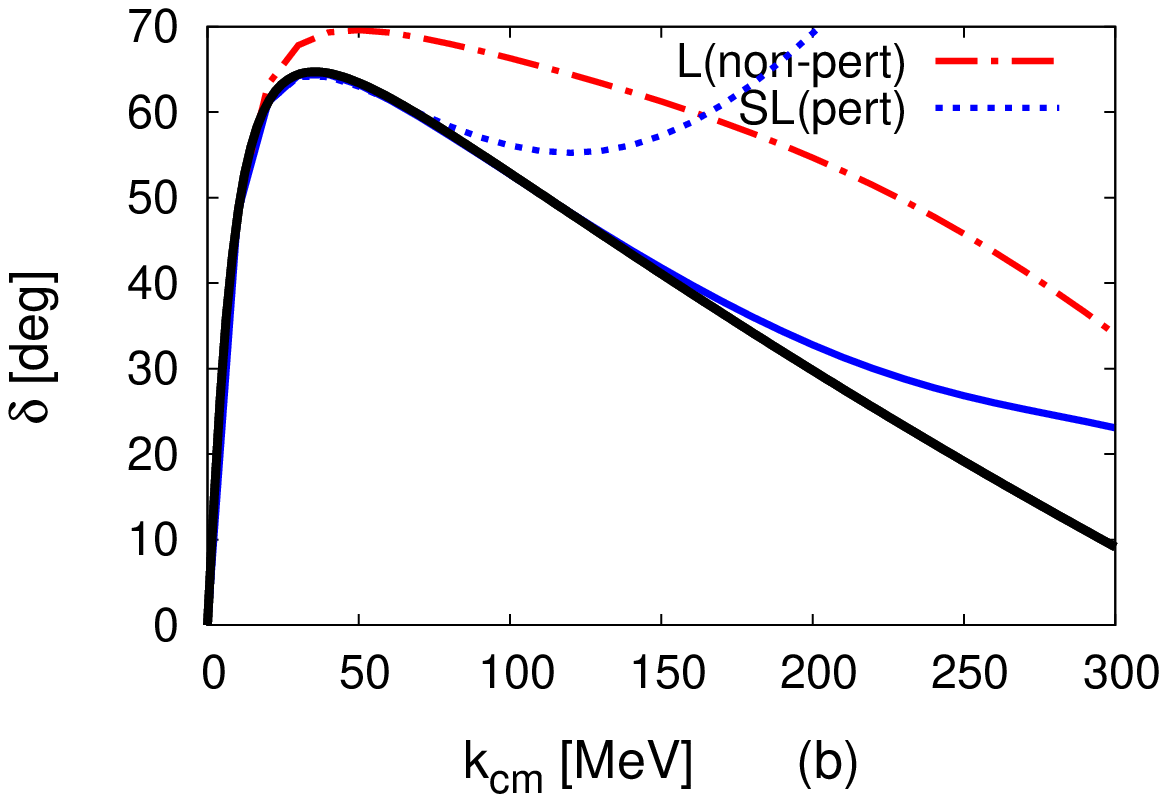}
\includegraphics[height=4.0cm, width=5.0cm]{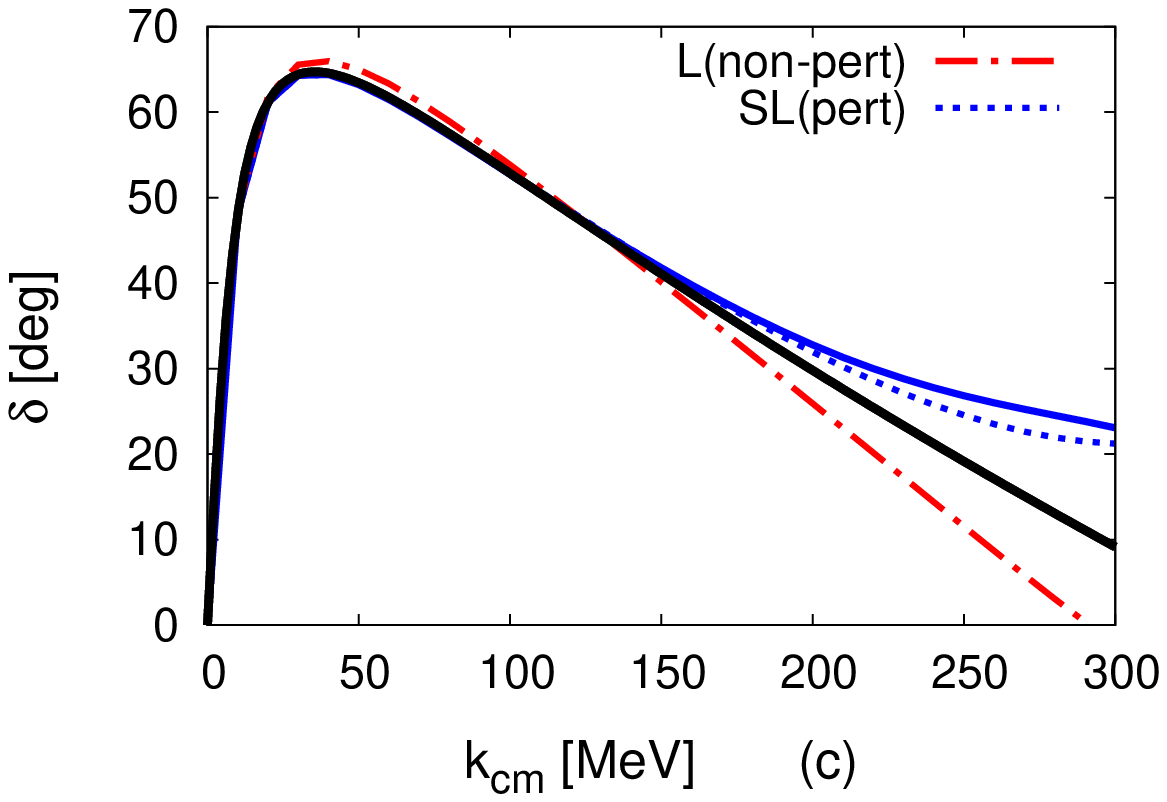}
\caption{
In panel (a), the phase shifts for the $^1S_0$ singlet channel computed
in the (non-perturbative) Weinberg counting at ${\rm N^2LO}$ 
by fitting the counterterms to the phase shifts for $k \leq 200\,{\rm MeV}$.
(b) The ${\rm N^2LO}$ phase shifts are approximated by an scheme in which
the leading order (L) consists on OPE and the $C_0$ counterterm,
while in the subleading order (SL) calculation TPE and the $C_2$ counterterm
are added in perturbation theory. (c) In this case the ${\rm N^2LO}$
phase shifts are approximated at leading order by TPE and $C_0$ and
at subleading order (perturbative) OPE and $C_2$.
}
\label{fig:Weinberg-breakdown}
\end{figure}

The previous question can be answered by playing different games with
the $^1S_0$ phase shifts of Fig.~\ref{fig:Weinberg-breakdown}.
From power counting we expect that the full T-matrix can be approximated
by a T-matrix in which the leading piece of the interaction has been
iterated and the subleading pieces have been included as perturbations:
\begin{eqnarray}
T(k) = \underbrace{T^{(0)}(k)}_{\mbox{Non-perturbative}} 
\quad + \quad
\underbrace{T^{(2)}(k) \, + \, T^{(3)}(k)}_{\mbox{Perturbative}}
\quad + \quad {\mathcal{O}}({Q^4}/{\Lambda_0^4}) \, .
\end{eqnarray}
This particular scheme is fulfilled in the panel (b) of
Fig.~\ref{fig:Weinberg-breakdown},
where we can see that power counting fails already at $k \simeq 100\,{\rm MeV}$.
Of course, this is an unsatisfactory situation, but it can get worse.
In fact we can try the following {\it anti} power counting scheme,
\begin{eqnarray}
T(k) = \underbrace{T^{(0)}(k)}_{\mbox{Perturbative}} 
\quad + \quad
\underbrace{T^{(2)}(k) \, + \, T^{(3)}(k)}_{\mbox{Non-perturbative}}
\quad + \quad {\mathcal{O}}({Q^4}/{\Lambda_0^4})\,
\end{eqnarray}
in which the subleading pieces of the interaction behave as
leading order pieces.
This power counting is realized in panel (c) of
Fig.~\ref{fig:Weinberg-breakdown},
and results in a very good approximation to the full phase shifts.
That is, the previous proposal is the underlying power counting of the
specific ${\rm N^2LO}$ calculation of Fig.~\ref{fig:Weinberg-breakdown}~\footnote
{However, in most cases it is impossible to uncover any underlying
power counting scheme.}.
This power counting {\it extravaganza } is a nice example of the kind
of unexpected behaviours discussed by Lepage~\cite{Lepage:1997cs}:
an order $Q^3$ operator is behaving as being of order $Q^{-1}$.

\section{Perturbative Renormalizability of Chiral TPE}

\subsection{Perturbative Weinberg Counting}

A particular way to avoid these power counting inconsistencies is
to treat the subleading order interactions in perturbation theory.
If subleading contributions to the amplitude are expected to be small,
it is natural to treat this pieces perturbatively.
This choice, which we will call perturbative Weinberg,
enforces power counting in observables
independently of the cut-off.
However, there are still problems with hard enough cut-offs,
like poor convergence of the chiral expansion or divergences.

Perturbative Weinberg has been explored by Shukla et al.~\cite{Shukla:2008sp}
for the particular case of the singlet $^1S_0$ channel,
where they find that the short range physics is compatible
with Weinberg in the range of coordinate space cut-offs
$r_c = 1.4-1.8\,{\rm fm}$ (but fails for $r_c < 1.0\,{\rm fm}$).
The recent lattice EFT calculations of
Refs.~\cite{Borasoy:2006qn,Borasoy:2007vi,Epelbaum:2009zsa,Epelbaum:2009pd,Epelbaum:2010xt}
(see also D. Lee's contribution to these proceedings)
also treat the subleading operators perturbatively.
They employ a spatial (temporal) lattice spacing of $a_s = 1.97\,{\rm fm}$
($a_t = 1.32\,{\rm fm}$) and provide an acceptable description of NN
phase shifts for $k \leq 100\,{\rm MeV}$~\cite{Borasoy:2007vi}
and of light nuclei~\cite{Epelbaum:2009pd,Epelbaum:2010xt}.
These results are encouraging for any perturbative setup;
however, they are limited to rather soft cut-offs for good reasons.

\subsection{Renormalizability and Modifications to the Power Counting}

\begin{figure}
\includegraphics[height=5.0cm, width=7.0cm]{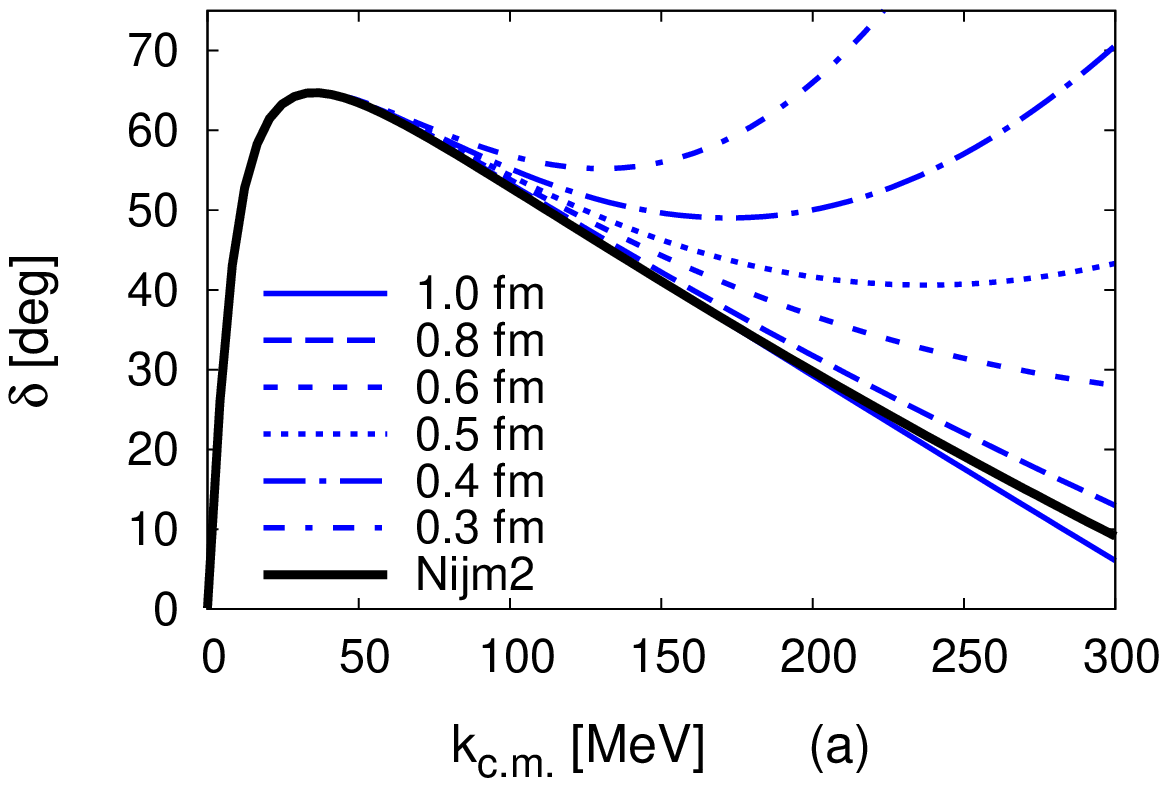}
\includegraphics[height=5.0cm, width=7.0cm]{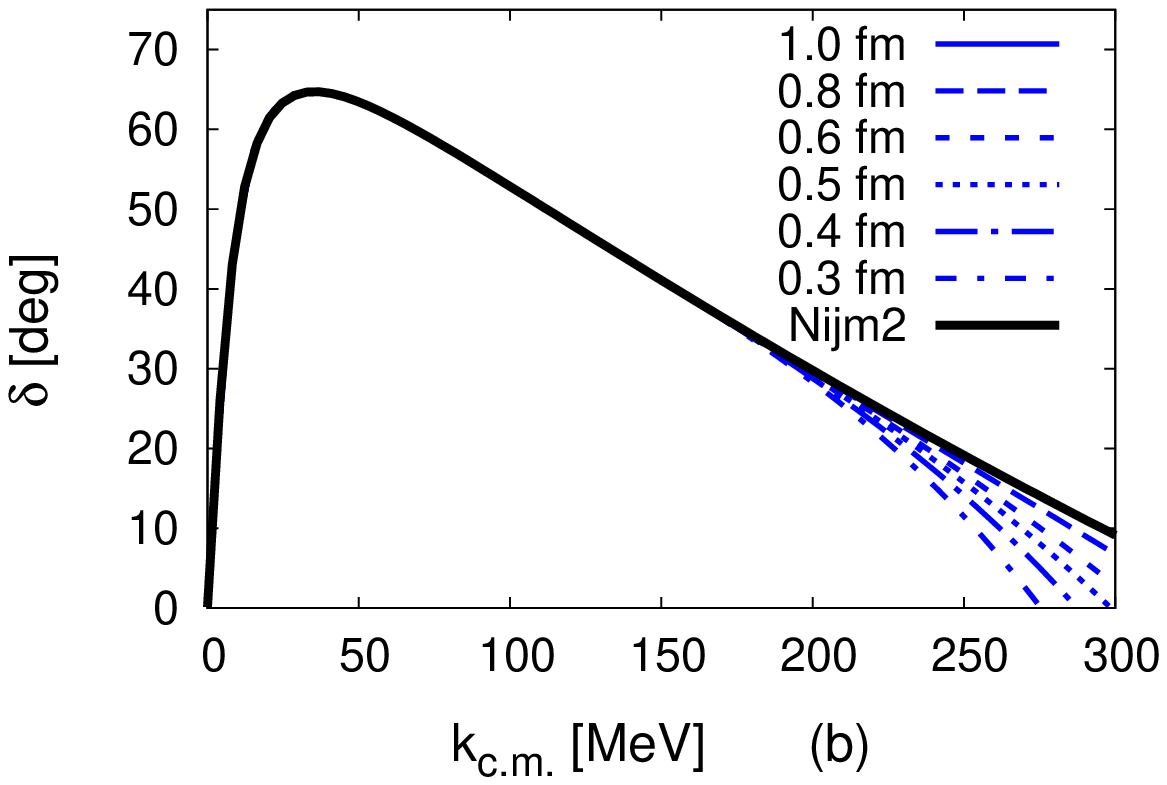}
\caption{
Phase shifts for the $^1S_0$ singlet channel with non-perturbative OPE
and perturbative TPE at ${\rm N^2LO}$.
The left panel (a) shows the resulting phase shifts
when only the counterterms prescribed by Weinberg counting
(i.e. $C_0$ and $C_2$) are included in the calculations.
In such a case the ${\mathcal O}(Q^3)$ contribution to the phase shifts
diverges as $k^4/r_c$, being $k$ the center of mass momentum and $r_c$
the coordinate space cut-off.
The right panel (b) shows the ${\rm N^2LO}$ phase shifts when the additional
$C_4$ counterterm is included.
The residual cut-off dependence of the phase shifts is now proportional to
$k^6\,r_c$.
}
\label{fig:cut-off}
\end{figure}

The problem with perturbative Weinberg is that the scattering amplitude
diverges for hard cut-offs.
In the particular case of the singlet $^1S_0$ channel,
we find that only using the counterterms prescribed by Weinberg,
that is 
$V^{(\nu = 2,3)}_{\chi,\rm contact} = C_0^{(\nu)} + C_2^{(\nu)} (p^2 + p'^2)$,
leads to the following divergences in the subleading pieces of the T-matrix:
\begin{eqnarray}
T(\Lambda) = T^{(0)}(\Lambda) + \underbrace{T^{(2)}(\Lambda)}_{\sim \log{\Lambda}} + \underbrace{T^{(3)}(\Lambda)}_{\sim \Lambda} +
{\mathcal{O}}({Q^4}/{\Lambda_0^4})\, .
\end{eqnarray}
In panel (a) of Fig.~\ref{fig:cut-off} we can see this linear divergence of
the ${\rm N^2LO}$ (order $Q^3$) phase shifts for a coordinate space
computation (the relation between the coordinate and momentum space
cut-off is roughly $r_c \simeq \pi / 2 \Lambda$~\cite{Entem:2007jg}).
The previous divergences can be cured by adding a new counterterm
at ${\rm NLO}$, that is, by taking
$V^{(\nu = 2,3)}_{\chi,\rm contact} = C_0^{(\nu)} + C_2^{(\nu)} (p^2 + p'^2) +
C_4^{(\nu)} (p^4 + p'^4)$.
This is equivalent to modify the power counting rules for the $C_4$ operator,
which is promoted from order $Q^4$ to order $Q^2$,
as determined by Birse~\cite{Birse:2005um}.
In panel (b) of Fig.~\ref{fig:cut-off} we can see the ${\rm N^2LO}$
phase shifts when the $C_4$ operator is included
in the computations,
leading to an amplitude which is free of ultraviolet divergences.

Of course this is not new:
five years ago, Nogga, Timmermans and van Kolck~\cite{Nogga:2005hy}
discovered by the numerical exploration of a large range of cut-off values
that the renormalizability of the Lippmann-Schwinger equation for
the leading order OPE potential required certain modifications
to the Weinberg counting.
This possibility was in fact anticipated in Ref.~\cite{PavonValderrama:2005gu}
on purely analytical grounds and can be easily understood in terms of
the non-perturbative renormalization of singular interactions~\cite{Valderrama:2005wv,PavonValderrama:2007nu,Entem:2007jg,PavonValderrama:2010fb}.
In Ref.~\cite{Birse:2005um} Birse studied in detail the power counting
resulting from iterating OPE using renormalization group analysis (RGA)
techniques (see also Ref.~\cite{Birse:2009my} for a more informal exposition).
This power counting is compatible with the short range physics extracted
in Refs.~\cite{Birse:2007sx,Birse:2010jr,Ipson:2010ah}
from the {\it deconstruction} of the phenomenological phase shifts
with perturbative TPE.

Recently, in Ref.~\cite{Valderrama:2009ei} the ${\rm NLO}$ and ${\rm N^2LO}$
phase shifts for central waves in the NTvK counting
have been obtained for the first time,
resulting in a good description of both the $^1S_0$ singlet and
the $^3S_1-{}^3D_1$ triplet phase shifts.
The power counting is determined by requiring perturbative renormalizability,
as explained previously for the particular case of the singlet channel.
The counting obtained in this way is basically equivalent to the one proposed
by Birse~\cite{Birse:2005um}, with some minor differences
in the triplet channel~\footnote{This may be a consequence of
the simplifications made in Ref.~\cite{Birse:2005um}
for treating the coupled channels}.

\begin{figure}
\includegraphics[height=5.0cm, width=7.0cm]{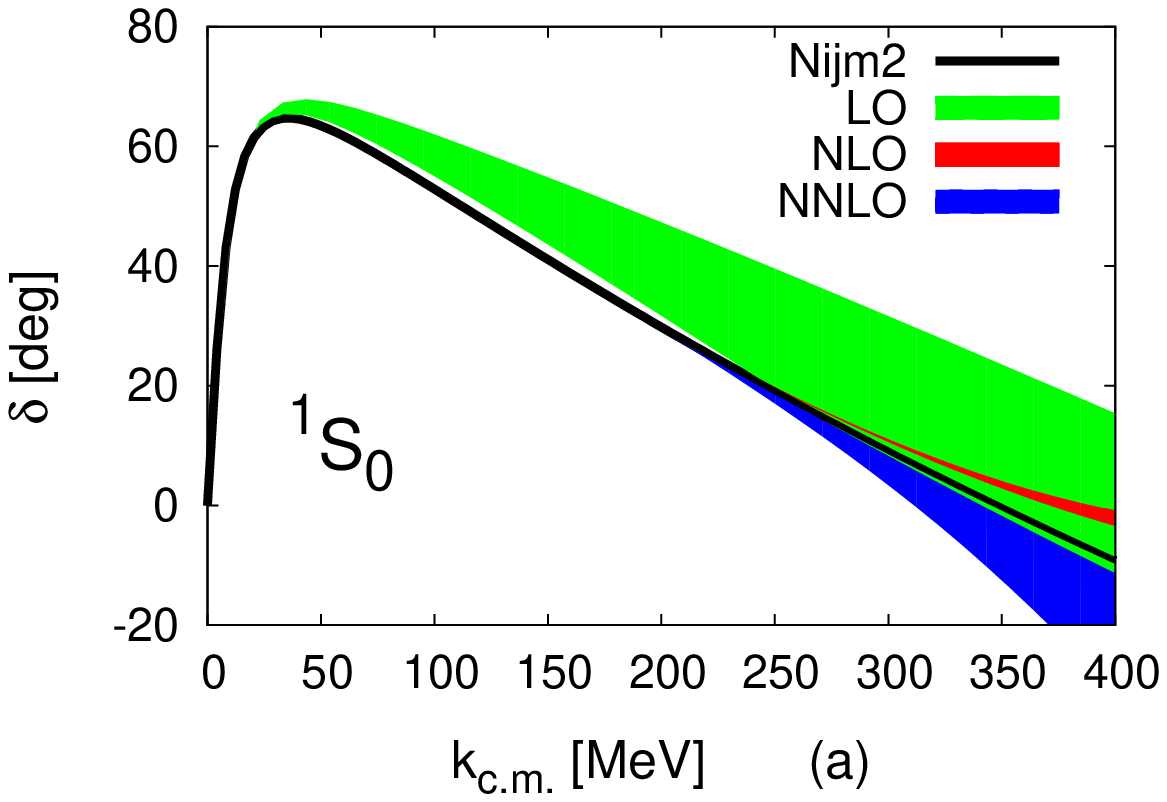}
\includegraphics[height=5.0cm, width=7.0cm]{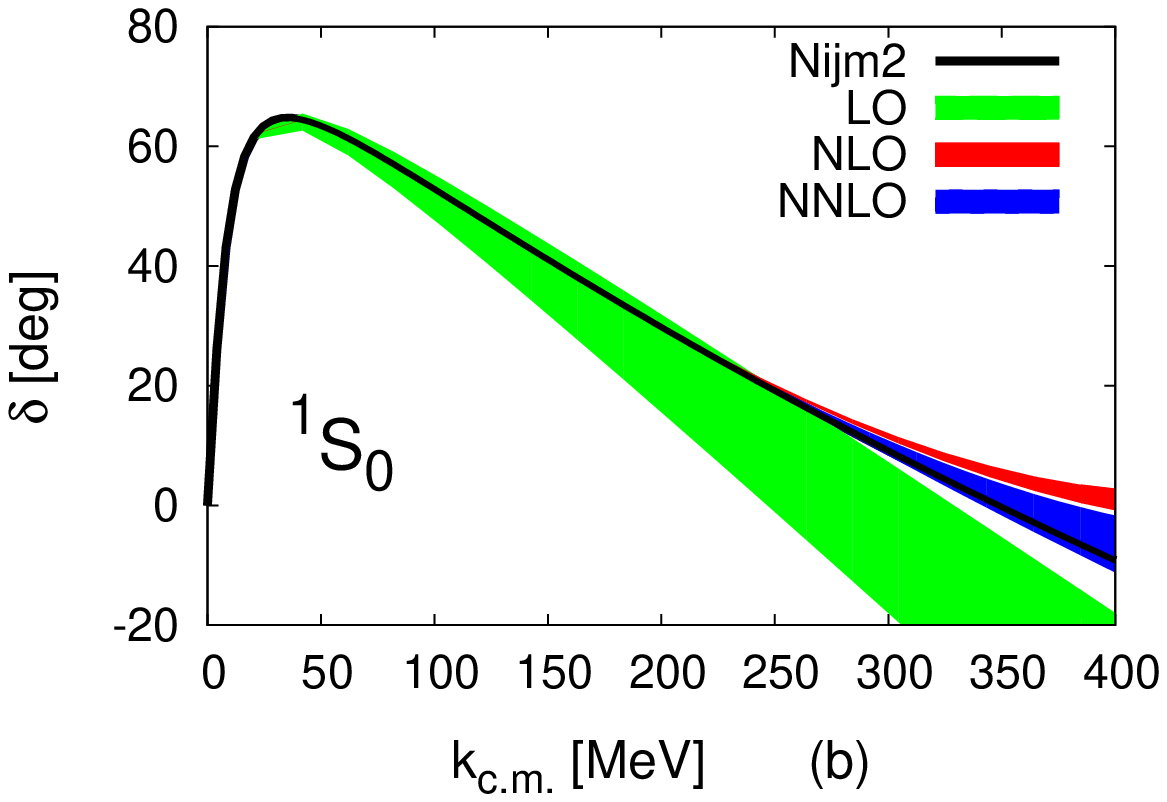}
\caption{
Phase shifts for the $^1S_0$ singlet channel with non-perturbative OPE
and perturbative TPE at ${\rm LO}$, ${\rm NLO}$ and ${\rm N^2LO}$.
The left panel (a) shows the resulting phase shifts in the Nogga, Timmermans,
van Kolck counting for the cut-off range $r_c = 0.6-0.9\,{\rm fm}$.
The right panel (b) shows the phase shifts for the cut-off
range $r_c = 0.9-1.2\,{\rm fm}$.
}
\label{fig:NNLO}
\end{figure}

In Fig.~\ref{fig:NNLO} we can see the ${\rm N^2LO}$ phase shifts
in the singlet channel for the cut-off ranges $r_c = 0.6-0.9\,{\rm fm}$
and $r_c = 0.9-1.2\,{\rm fm}$.
In panel (a) we follow the philosophy of Ref.~\cite{Epelbaum:2006pt},
in which the cut-off is varied around the purported breakdown scale,
probably $0.7-0.8\,{\rm fm}$, the distance at which
most phenomenological potentials have their minima
(signalling the point in which the short range repulsion starts
to become stronger than the long range attraction).
In panel (b) we follow Ref.~\cite{Beane:2008bt}, in which the cut-off is
interpreted as a parameter controlling the convergence of the EFT
expansion: a softer cut-off in general improves convergence.
However, in both cases the phase shifts are pretty similar.

\subsection{Which is the correct value for the Cut-off?}

The correct value of the cut-off has become a contentious issue
in nuclear EFT.
Here we will first consider the problem of the cut-off
from the point of view of the practical advantages
that certain cut-off values entail.
Then we will propose a possible interpretation.

Perturbative renormalization guarantees that the amplitudes are free
of divergences when the cut-off is removed.
However, there is still a serious problem with removing the cut-off:
the convergence of the perturbation series is not assured
for cut-off radii below $r_c \simeq 0.6-0.7\,{\rm fm}$.
This is due to the appearance of the first deeply bound state,
which generates a pole in the renormalization group evolution
of the $C_0(r_c)$ counterterm,
a feature that cannot be reproduced perturbatively.
As a consequence, the related expansion for $C_0(r_c) =
C_0^{(0)}(r_c) + C_0^{(2)}(r_c) + C_0^{(3)}(r_c) + \mathcal{O}(Q^4)$
must fail for $r_c < 0.6-0.7\,{\rm fm}$, spoiling the power counting.

On the other hand, if the cut-off is of the order of $1.4-1.8\,{\rm fm}$,
there is no practical advantage on modifying the power counting,
at least in the singlet, as the phase shifts are already
well described by the perturbative Weinberg setup of Ref.~\cite{Shukla:2008sp}.

In this regard, not all regularization schemes or cut-off values are adequate
to realize a particular power counting proposal.
Probably the best example is
the KSW counting~\cite{Kaplan:1998tg,Kaplan:1998we},
which can be implemented with power divergence subtraction (PDS)
but not with minimal subtraction (MS).
As was shown by Cohen and Hansen~\cite{Cohen:1998bv},
KSW can also be realized as a cut-off theory
(see also the related observations of Ref.~\cite{Epelbaum:2009sd}).
However, independently of whether dimensional or cut-off regularization is used,
there are certain limitations on the regularization scale.
The analysis of the running of the $C_0$ counterterm
of Ref.~\cite{Kaplan:1998we}
requires $\mu < \Lambda_{\rm NN} \sim 300\,{\rm MeV}$,
or equivalently $r_c > 1 / \Lambda_{\rm NN}$ if the arguments
are applied to a coordinate space cut-off,
for the KSW counting to work.

From the previous we are tempted to interpret that the role of regularization
and renormalization is to guarantee that power counting
is correctly implemented.
In this regard, the main inconsistency in current implementations
of Weinberg is the power counting {\it extravaganza}
phenomenon discussed earlier, which is in fact a consequence
of the cut-off range employed in the calculations.
This interpretation is complementary to the role suggested for the cut-off
by Beane et al.~\cite{Beane:2008bt},
in which the cut-off may be chosen in order to improve the convergence
of the expansion.
However, as explained by Birse~\cite{Birse:2010jr}, these interpretations
are unorthodox and require a firmer theoretical basis.
In particular, for perturbative chiral TPE the optimal cut-off range lies
probably in the vicinity of $\sim 1\,{\rm fm}$ or above.

\section{Conclusions}

The perturbative treatment of chiral two pion exchange provides
the opportunity to construct scattering amplitudes with are
compatible with the requirements of renormalizability
and power counting within an effective
field theory framework~\cite{Valderrama:2009ei}. 
On the contrary, non-perturbative approaches are not guaranteed
to fulfill these conditions; in particular, power counting may
be lost if the cut-off is too hard, as exemplified with an
example Weinberg calculation at ${\rm N^2LO}$.
The perturbative TPE calculations show the feasibility of the Nogga,
Timmermans and van Kolck proposal~\cite{Nogga:2005hy}.
While the role of the cut-off is still (and will be for some time)
a controversial issue, we propose a sensible interpretation
in the line of Ref.~\cite{Beane:2008bt} which may be able to
reconcile the difficult requirements for the EFT formulation
of a problem with a poor separation of scales.


\begin{theacknowledgments}
I would like to thank E.~Ruiz Arriola for discussions and
comments on this manuscript and E.~Epelbaum for discussions.
This work is supported
by the Spanish Ingenio-Consolider 2010 Program CPAN (CSD2007-00042)
and by the EU Research Infrastructure Integrating Initiative HadronPhysics2. 
\end{theacknowledgments}

\bibliographystyle{aipproc}   


\end{document}